\documentstyle[referee]{laa}     

\begin{document}

   \thesaurus{ 11.03.1;  
               11.07.1;  
               12.07.1;  
             }
   \title{Galaxy-cluster associations from gravitational lensing}


   \author{Xiang-Ping Wu$^{1,2}$, Francois Hammer$^{1,3}$}

   \offprints{Xiang-Ping Wu$^1$}

   \institute{$^1$DAEC, Observatoire de Paris-Meudon, 92195 Meudon Principal
Cedex,
               France\\
              $^2$Beijing Astronomical Observatory, Chinese Academy of
Sciences,
                Beijing 100080, China\\
               $^3$CFHT, PO Box 1597, Kamuela, HI 96743, USA}

    \date{Received \qquad\qquad, 1993; accepted \qquad\qquad 1994}

   \maketitle

   \begin{abstract}
%
We investigate the associations between background
galaxies and foreground clusters of galaxies due to the
effect of gravitational lensing by clusters of galaxies.
Similar to the well-known quasar-galaxy ones,
these associations depend sensitively on the shape of galaxy
number-magnitude or number-flux
relation, and both positive and ``negative" associations are found to be
possible, depending on the limiting magnitude and/or the flux threshold
in the surveys.  We calculate the enhancement factors assuming a singular
isothermal sphere model for
clusters of galaxies and a pointlike model for background sources selected
in three different wavelengths, $B$, $K$ and radio.
Our results show that $K-$ selected galaxies might constitute the best sample
to test the ``negative" associations while it is unlikely that one can
actually observe any association for blue galaxies.
We also point out that bright radio sources ($S>1$ Jy) can provide
strong positive associations, which may have been
already detected in 3CR sample.
      \keywords{ gravitational lensing -- galaxies: general --
                 galaxies: clustering}
   \end{abstract}

%

\section{Introduction}

It has been argued for two decades about the existence of
associations between
high redshift quasars and low redshift galaxies. If these associations
are real but not physical, gravitational lensing has thus far provided
the most natural explanations. Indeed, quasar-galaxy associations had been
predicted (Gott \& Gunn, 1974)
even before the discovery of the first lensed quasar pair. It is generally
believed that distant quasars can be magnified by gravitational lensing of the
matter associated with foreground galaxies,
leading to an overdensity of quasars around the
galaxies.  Nevertheless, the present observational evidences on quasar-galaxy
associations appear to be puzzling: positive, null and even  negative
evidences have been reported in different observations (Narayan, 1992),
in contradiction with previous
claims based on gravitational lensing.  Wu (1993) has very recently noticed
that gravitational lensing can not only produce a number increase of
background quasars in the vicinities of foreground galaxies, but also can
reduce the quasar surface number density, depending sensitively  on
the ``turnover" in the number-magnitude relation and the limiting magnitude
of background quasars (Boyle, Shanks and Peterson, 1988).
For example,  one can expect
positive associations for bright quasar samples (limiting magnitude of
$B<19.5$)
 while ``negative" associations for fainter magnitude limits. The latter are
the
consequence of the flatter slope in  quasar counts.
Such scenarios have successfully explained the current
observations of quasar-galaxy associations.

The motivation of the present paper is then to extrapolate the quasar-galaxy
associations to some other larger systems, namely, the associations between
background galaxies and foreground clusters (hereafter galaxy-cluster
associations), and explore the observational possibilities.  Recall that
any background sources whose number-magnitude relations do not show the slope
of $\log N/dm= 0.4$ is potentially affected by gravitational lensing,
resulting in either positive or ``negative" associations with
foreground objects.

It was noticed a long time ago that there may exist
 associations between galaxies and galaxy clusters.
Roberts, O'Dell \& Burbidge (1977)  found
more nearby 3CR galaxies near the positions of clusters of galaxies
although their radio sources might be physically associated
with their host clusters.
The evidences on these associations have
remained unclear until recent years when
some of the high redshift 3CR galaxies are found to be gravitationally
magnified by foreground clusters of galaxies lying in their lines of sight
(Hammer, Nottale \& Le F\`evre, 1986; Le F\`evre, Hammer \& Jones, 1988;
Le F\`evre et al., 1988; Hammer \& Le F\`evre, 1990).
Indeed, bright radio galaxies are ideal sources to be lensed since they are
usually at high redshifts ($\langle{z}\rangle\approx1$--$2$)
and  the slope of their counts is extremely steep.
At optical wavelengths,  the Hubble
diagram of the brightest cluster galaxies might be also affected by foreground
clusters (Hammer and Nottale (1986)).
Moreover, the presence of giant luminous arcs in the cores of rich clusters
(Hoag, 1981; Lynds \& Petrosian, 1986; Soucail et al, 1987) is
a direct evidence that rich clusters might affect
the observations of faint background galaxies.

Galaxy-cluster associations depend on the selection of flux and wavelength of
background galaxies. In this paper we consider $B$, $K$ and radio selected
galaxies
and study their associations with foreground clusters.

%

\section{Number counts and enhancement factors of background galaxies}

The ratio of the observed surface number density of background sources
to their undisturbed
value is referred to as the enhancement factor $q$.  $q$ is balanced
by two factors: magnification ($A$) and area distortion ($1/A$).
The first effect leads to the
increase of total number in a flux-limited survey by picking up
fainter sources and the second one
reduces the surface number density by enlarging the searched area.
$q$ can be simply expressed as (Narayan, 1989)
\begin{equation}
q=\frac{N(<m+2.5\log A)}{N(<m)}\frac{1}{A}
=\frac{N(>S/A)}{N(>S)}\frac{1}{A}.
\end{equation}
Where $N$ is the number counts of background sources,
$m$ and $S$ denote
the apparent magnitude and flux, respectively.
Hence, $q$ can be determined by two parameters only:
the intrinsic or the undisturbed number counts of
background sources and the lensing magnification by foreground objects.
Leaving $A$ to be a free parameter, one can calculate the enhancements
for different background sources by assuming that the observed
number-magnitude or number-flux
relations have not been significantly contaminated by gravitational
lensing and can, thereby,  represent rather well their intrinsic ones.
Indeed, it is very unlikely that lensing can actually change the whole
behaviours of the total number counts of any kinds of sources [e.g., for
quasars see Schneider (1992)].

The cumulative surface number density of galaxies in $B$ and $K$ bands can be
 well fitted by the following expressions (Tyson, 1988; Cowie \& Songaila,
1992)
\begin{equation}
N(<B)=1.72\times 10^{4+0.45(B-24)}
\end{equation}
and
\begin{equation}
N(<K)=\left\{
\begin{array}{ll}
2.1\times 10^{3+0.63(K-17)}, & \;\;K<17;\\
-5.6\times10^3+7.8\times10^{3+0.26(K-17)}), &\;\;K>17.
\end{array}\right.
\end{equation}
For radio sources we use the number counts $N(S)$ of
Dunlop and Peacock (1990) and Langston et al. (1990) ( see also Wu \& Hammer,
1993).
An important feature of $N(S)$ is that
at bright end ($S>1$ Jy) the slope
   \begin{figure}
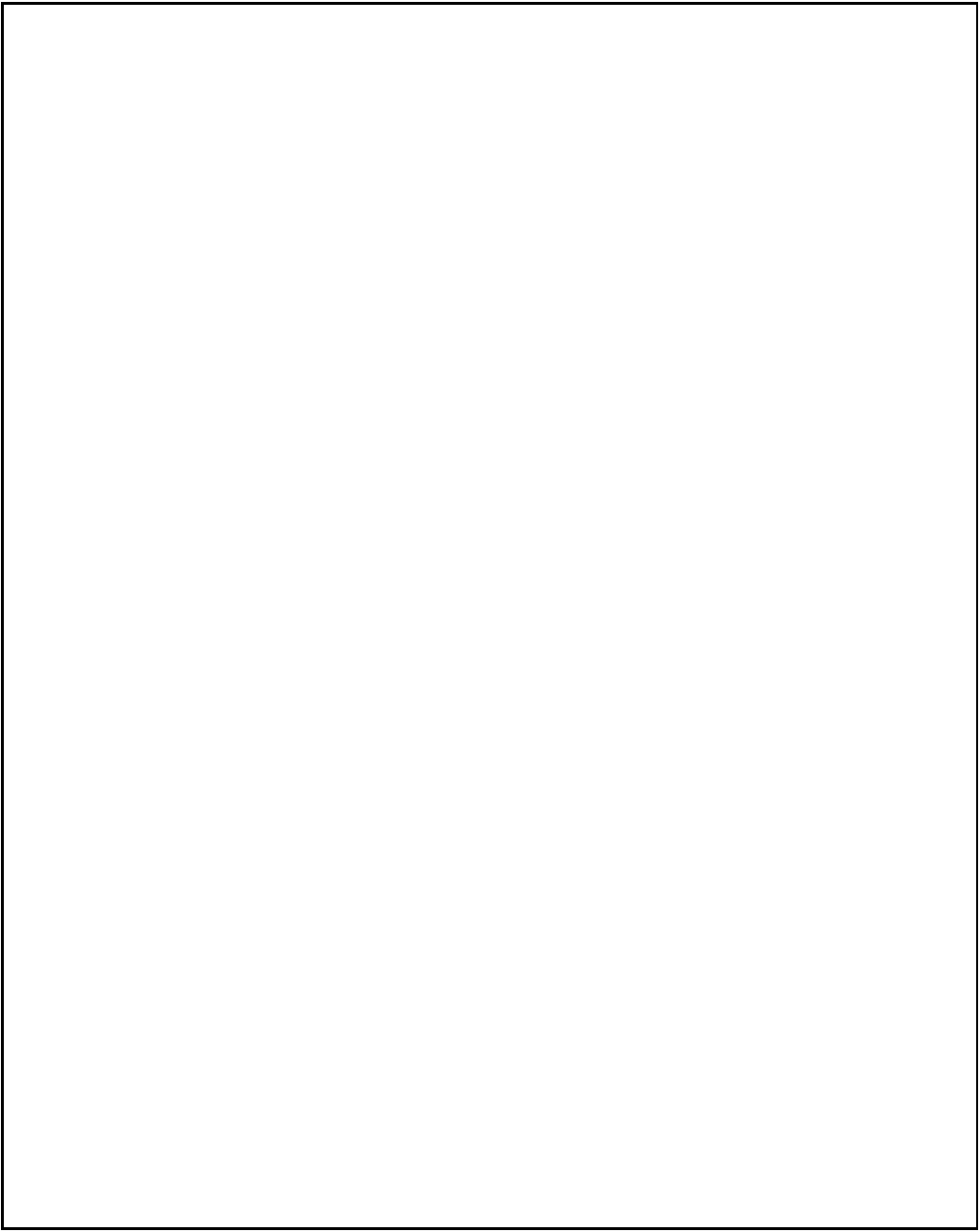

      \picplace{17cm}
      \caption{The enhancement factor $q$ against magnification
$A$ by gravitational lensing for galaxy-cluster associations.
(a)$B$-selected galaxies; (b)$K$-selected galaxies; (c)Radio sources.
	}
         \label{Fig.1}
    \end{figure}
turns to be very steep, hence one would expect such sources to be
significantly affected by lensing, while at faint end $N(S)$ is relatively
flat. However,
radio sources are not only galaxies but also quasars and
the fraction of quasars in radio source surveys varies with flux
threshold. The following discussion on
radio galaxy-cluster associations should be then generally considered to
include radio quasars although we mainly discuss galaxies as sources.

The enhancement factors against lensing magnification
for different galaxies can be calculated by
eq.(1), and their results are shown in Fig.1.  Because the $B-$ galaxy counts
have a constant slope just above 0.4, large enhancement factors of up to $2$
are
 unlikely for realistic lensing objects. Hence no significant associations of
blue
galaxies with clusters are expected.
Associations between clusters and
$K$- selected galaxies are very similar to quasar-galaxy associations
(Wu, 1993), and can be either positive and ``negative"
 because of the ``knee" of $K\sim17$ in the counts.
Bright radio sources
($S>1$ Jy) are the most promising objects for searching strong associations
with clusters (large $q$ values) while the enhancement of
fainter sources with $S<10$ mJy appears to be  negative ($q<1$).
Note, however, that  the above calculations require the
extrapolations of source number counts into fainter
magnitude/flux.  The subsequent quantities of $m+2.5\log A$ and $S/A$ in
the number counts of eq.(1) may exceed observational limiting
magnitude or flux threshold when $A$ becomes very large.

%

\section{Galaxy clusters as lenses}

Lensing properties may depend on {\it a priori} choice of the mass profiles
for lensing objects (Wu \& Hammer, 1993) whilst the real matter distribution
in clusters of galaxies is still poorly known.  It is commonly believed that
the isothermal model and the King model can both fit fairly well the luminous
matter (optical/X-ray) distribution in clusters of galaxies. However,
the presence
of giant arcs in rich distant clusters implies highly concentrated dark matter
in
their cores (Hammer, 1991; Wu \& Hammer, 1993; Le F\`evre et al., 1993).
For the purpose
of study of galaxy-cluster associations,  only moderate and small
magnifications
are actually concerned, i.e., the regions to be considered are far from the
central cores.  It is then convenient to assume a singular isothermal sphere
for matter distribution in clusters, which would greatly simplify
the calculations.

For simplicity we consider a pointlike source for the background galaxy
and leave the discussion about
extended sources to the next section.  The magnification
of a point source by a singular isothermal sphere  is
\begin{equation}
A=\frac{\theta}{\theta-\theta_c}.
\end{equation}
Where $\theta$ measures the distance to the center of a galaxy cluster with a
critical radius  (Einstein radius) of
\begin{equation}
\theta_c=4\pi\left(\frac{\sigma}{c}\right)^2\frac{D_{ds}}{D_s}.
\end{equation}
Here $D_{ds}$ and $D_s$ are the angular diameter distances from the lensing
cluster and from the observer to the distant sources, respectively.

Without the detailed knowledge of velocity dispersion in
clusters of galaxies, the typical
features of $q$ with the limiting magnitude/flux threshold can be found
by using the critical radius ($\theta_c$) as distance unit.
   \begin{figure}
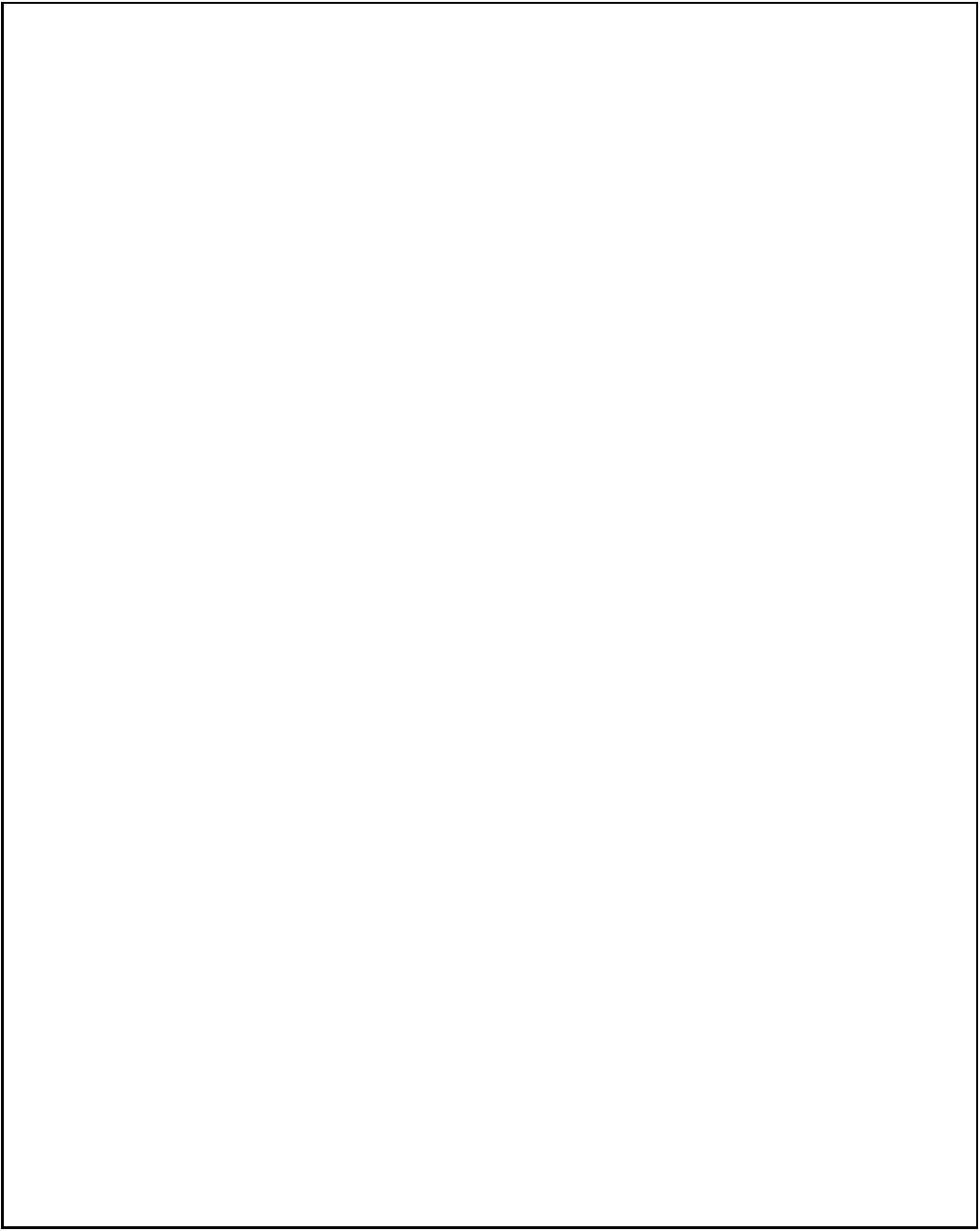

      \picplace{17cm}
      \caption{
The dependence of enhancement factor $q$ on
limiting magnitude/flux threshold. The search regions are chosen in
the unit of the critical radius ($\theta_c$).
(a)$B$-selected galaxies; (b)$K$-selected galaxies; (c)Radio sources.
}
         \label{Fig.2}
    \end{figure}
Fig.2 illustrates
the enhancement factors for $B$-, $K$- and radio selected galaxies
within the regions of 1.5, 2, 4 and 8 critical radii.  As it is expected,
blue galaxy-cluster associations are independent of limiting magnitude.
It is very likely that a null evidence on such associations way be reached
because of the corresponding low value of $q$ ($<1.1$).  Conversely to this,
infrared galaxies turn to present noticeably  both positive and ``negative"
associations.  For $K<17$ galaxies, a high value of $q$ is expected,
especially when one looks at the regions near the critical radius.
At the fainter end  ($K>18$), a considerably smaller number of infrared
galaxies would be found, leading to ``negative" associations. This
``negative" effect is even much stronger than in the quasar-galaxy
associations (Wu, 1993). The turnover point appears between $K=17$ and
$K=18$ because of the break of number counts at $K\approx17$.
It is then suited to choose a survey
limiting magnitude well outside this range (below or above) in order to
study $K$ galaxy-cluster
associations.  For radio sources, it is very hopeful to observe
 large enhancements at high flux limits. Choosing a flux threshold
of $S_{5{\rm GHz}}=3$--$4$ Jy would provide relatively larger values of
$q$.  Although the ``negative" associations exist for $S_{5{\rm GHz}}<30$ mJy,
it is unlikely that one can easily see them nowadays without considerably
large radio telescope time.

For a further discussion of clusters of galaxies as lensing objects,
one needs to  consider two
parameters appearing in the critical radius $\theta_c$: the distance factor
$D_{ds}/D_s$ and the velocity dispersion $\sigma$.
If one can eliminate the factor of
$D_{ds}/D_s$ in eq.(5), the uncertainties in distances to foreground clusters
and background galaxies would vanish in the calculation of enhancement factor
$q$.  This can be indeed reached by properly choosing nearby clusters of
galaxies and distant sources so that $D_{ds}/D_s\approx1$.
Nevertheless, the distance
parameter $D_{ds}/D_s$ only affects the critical radius.
Therefore, taking into account of the actual distances of the clusters and the
galaxies is equivalent to having a lens with smaller critical radius, which
does
not change the $q$ in shape shown in Fig.2.
The following calculations will then be made
for the ``maximum" Einstein radius assuming $D_{ds}/D_s=1$.

   \begin{figure}
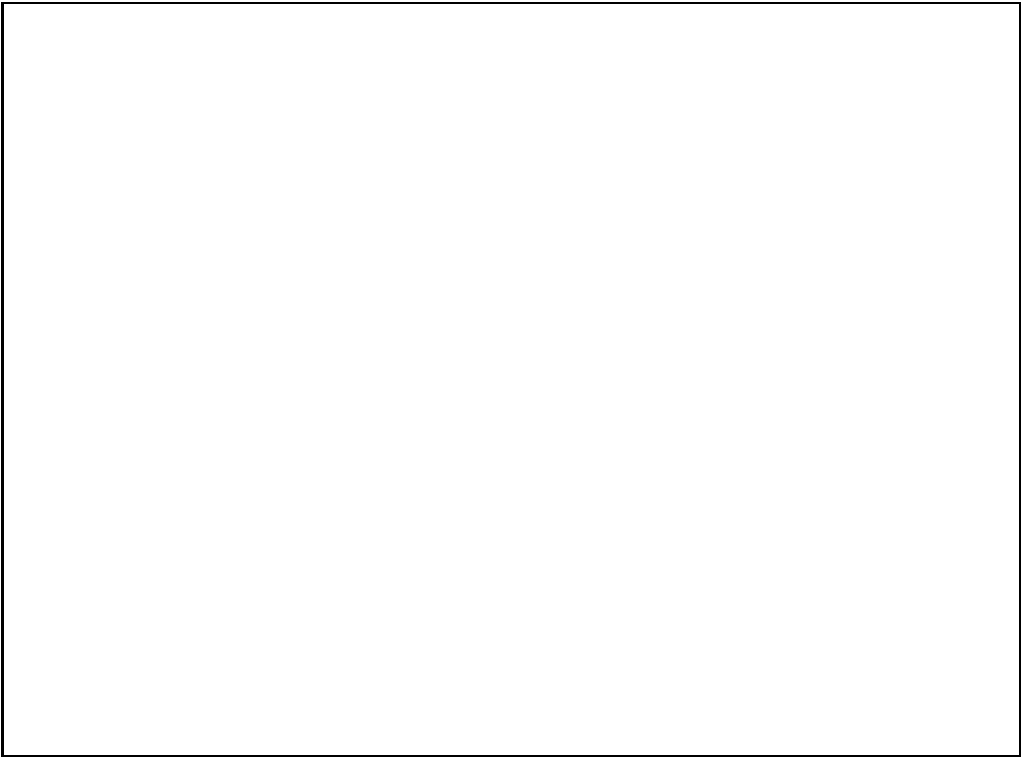

      \picplace{10cm}
      \caption{
The enhancement factors $q$ of $B$-selected galaxies versus
search ranges ($\theta$) in nearby poor ($\sigma=600$ km/s)
and very rich ($\sigma=1200$ km/s) galaxy clusters. }
         \label{Fig.3}
    \end{figure}
   \begin{figure}
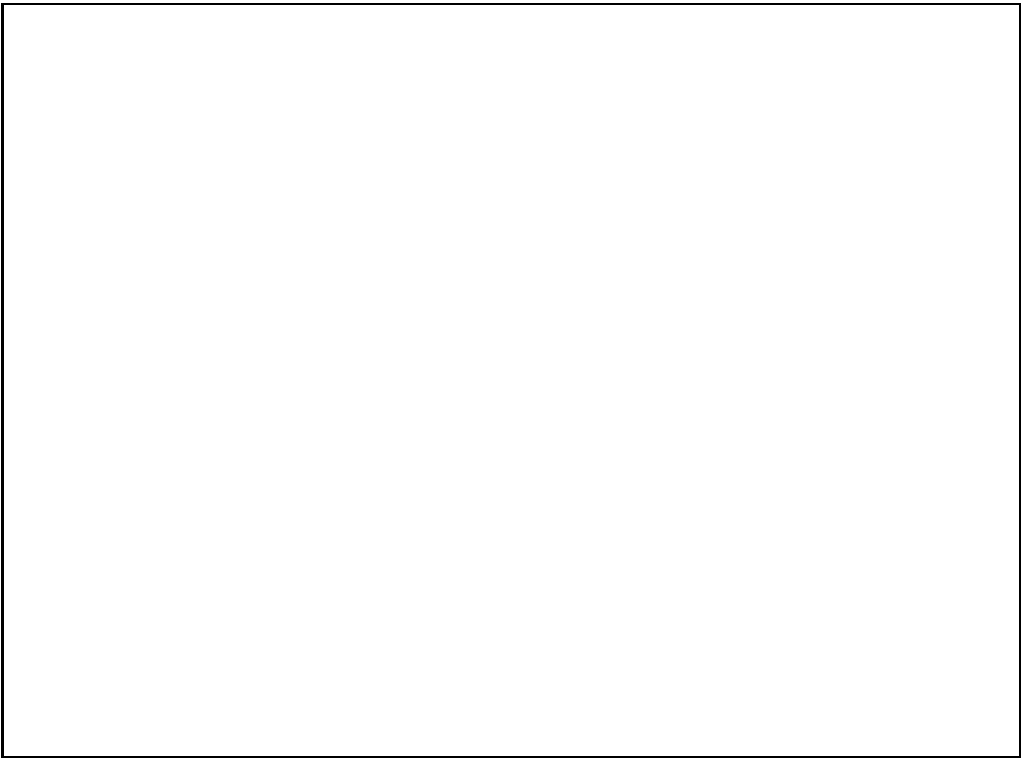

      \picplace{10cm}
      \caption{
The same as Fig.3 but for $K$-selected galaxies.}
         \label{Fig.4}
    \end{figure}
   \begin{figure}
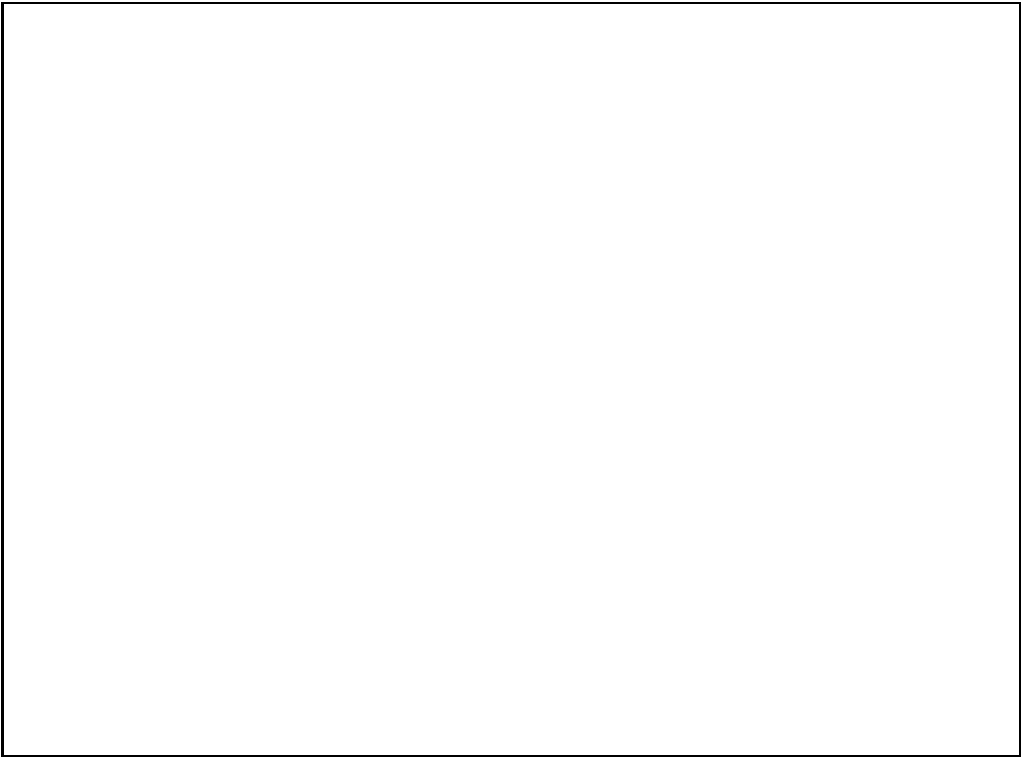

      \picplace{10cm}
      \caption{
The same as Fig.3 but for radio sources.}
         \label{Fig.5}
    \end{figure}
An exact treatment of finding $q$ for background galaxies would be made by
considering the luminosity ($L$) distribution of clusters of galaxies,
then to estimate their corresponding velocity dispersions ($\sigma$) throughout
a {\it quasi-} Faber-Jackson relation and finally to integrate over the
redshift space.
This procedure of statistical lensing by clusters of galaxies has been
extensively investigated by Wu and Hammer (1993). Nevertheless, statistical
lensing by clusters of galaxies may still have some uncertainties today,
mainly resulting from the fact that the $L-\sigma$ relation for clusters of
galaxies has not been very well determined. Here we simply consider
two individual clusters, one with
large $\sigma$ (rich cluster) and another with low $\sigma$ (poor cluster).
Fig.3 -- 5 give the results of $q$ derived from two nearby clusters with
$\sigma=1200$ km/s and $\sigma=600$ km/s, respectively,
for the three wavelength selected
galaxies, assuming that redshift of the sources is much larger than
redshift of the lenses.
 Compared to Fig.2, the similar features except the
amplitudes are shown. It strengthens that
nearby and very rich clusters of galaxies are the best
targets to test the galaxy-cluster associations.

%

\section{Discussion}

\subsection{Extended sources}

The above calculations have been made under the assumption that
background galaxies are pointlike. We now discuss if the extended sources
would contribute any significant effects on the $q$ derived from the
pointlike sources.  Indeed at small virtual impact parameters, magnification
factor reaches a maximum value for an uniform luminous disk (formation of
an Einstein ring) instead of a linear increase of magnification for a point
source.
This type of magnification distribution
is valid for any spherical lenses, and
such an example for a point lens can be found in Bontz (1979). Hence,
lensing magnification for an extended source by a spherical lens
can be approximated by using the magnification for a pointlike source till
the point where magnification exceeds its maximum value for a uniform
disk source.

For a singular isothermal sphere as deflector, the maximum magnification
of a uniform disk with radius of $R$ is
\begin{equation}
A_{max}=1+8\pi\left(\frac{\sigma}{c}\right)^2\frac{D_{ds}}{R}.
\end{equation}
Taking a standard cosmological scenario of $\Omega=1$ and $H_0=50$ km/s/Mpc,
eq.(6) reads
\begin{equation}
A_{max}=1+116.35\frac{\sqrt{1+z}-1}{(1+z)^{3/2}}
\left(\frac{\theta_c}{10''}\right)
\left(\frac{R}{10{\rm kpc}}\right)^{-1}
h_{50}^{-1}.
\end{equation}
A simple numerical calculation shows that for galaxies having
redshifts ranging from 0.2 to 3, $A_{max}$ is not smaller than
$\sim10$ even if $\theta_c$ is taken to be $10"$ (Note that this
critical radius corresponds to a poor galaxy cluster with $\sigma=600$ km/s,
see Fig.3).  Indeed, this maximum magnification
is large enough for the consideration of associations. Recall that only giant
 arc events can reach such a high magnification of $\sim 10$. Therefore,
if one does not choose the regions very close to the critical line,
the pointlike model for background galaxy would be a good approximation.
For instance, using $\theta=1.5\theta_c$ in Fig.2 gives $A=3$ and using
$\theta=50"$ in Fig.3 with $\theta_c=41"$ gives $A\approx6$, both of which
are smaller than the maximum magnification of $\sim10$  provided
that the source was an extended
galaxy with a luminous disk of $10$ kpc. Indeed, disk
radii of galaxies are usually smaller than $10$ kpc, implying that for moderate
magnification the point source hypothesis is reasonable.

\subsection{Optical galaxies}

As were shown in Fig.1--3, blue galaxy-cluster associations are unlikely
detectable because the slope of the blue galaxy counts is very close to 0.4.
Moreover, the relative excess of faint blue galaxies, provided by the fact that
there is
no apparent turn-over in their number-magnitude relation, is probably due to
an excess of dwarves at low redshifts (Cowie, Songaila \& Hu, 1992;
Tresse et al, 1993), the latter
being not the ideal sources for studying galaxy-cluster associations.

$K-$ selected galaxies span rather a larger range of redshifts.
The median redshift of galaxies is $0.6$ for $K=18$--$19$ (Cowie \& Songaila,
1992),
appearing to be the good targets for studying galaxy-cluster
associations.  Taking $K=19.5$ to be the limiting magnitude in the survey
would result in a significant effect of $q<1$ (see Fig.4), which might
provide a better example than in the quasar-galaxy surveys
to test the ``negative" associations.  Certainly, brighter
galaxies in $K$ can be also used to see the strong positive associations.
However,
contamination by cluster galaxies (especially ellipticals which are
concentrated
towards cluster centers) would render a difficulty for
a simple analysis based on images, implying that a dedicated (and telescope
time consuming) spectroscopic work would be required.

\subsection{Radio sources}

Bright radio sources provide the largest enhancement due to their steep
slope of number counts at $S>1$ Jy, constituting the ideal sources to
find enhancement factor $q$ for galaxy-cluster associations by
gravitational lensing. Indeed, five out the 23 3CR radio galaxies at
$z\geq1$ have been found to having foreground clusters of galaxies
within about 1 arcminute near the lines of sight (Hammer \& Le F\`evre,
1990), while the probability of observing these pairs randomly in the sky
is only $0.05\sim0.06$. It then remains very promising to soon have
a more statistically significant sample of radio galaxy-cluster
associations which can be selected from the known bright radio source surveys.
Thus, there might be no need for further radio observations
for such a purpose and the work of extracting $q$ in the catalogues of
bright radio source and clusters of galaxies is underway.

%

\section{Conclusions}

We have extrapolated the quasar-galaxy associations into larger
systems, namely galaxy-cluster associations, which result from gravitational
lensing
by foreground clusters of galaxies. We found that these associations
are indeed observable, especially for $K-$ and radio selected galaxies.
Similar to quasar-galaxy associations, enhancement factor
describing galaxy-cluster associations depends sensitively on
number-magnitude or number-flux relations of background galaxies. We have
pointed out the importance of the choice of the limiting magnitude or of the
flux threshold in the surveys, which relate closely to the enhancement
factors whether they are larger than unity (positive associations) or
smaller than unity (``negative" associations).  $K-$ selected galaxies
turn to be the good targets to test ``negative" associations" while
the radio sources would provide the strongest positive ones, already
detectable in the bright radio samples like 3CR.

To easily confirm galaxy-cluster associations, one needs to choose
nearby rich clusters of galaxies in the sense that the distance
$D_{ds}$ should be very close to $D_s$. This may help to provide large
Einstein radius, leading to large areas for searching
associated background galaxies enough far away from the central
core of galaxy cluster.
Rich clusters have large Einstein radius and provide significant
enhancement factors. Therefore, nearby Abell clusters with
$\sigma>1000$ km/s would be strongly recommended for the statistical study of
galaxy-cluster associations. A direct knowledge of velocity dispersion in the
cluster of galaxies would be also very useful to estimate in advance the
Einstein radius and then the theoretical enhancement factor.

Yet, the present paper deals with only the simple models for
matter distribution in and spatial distribution for clusters of galaxies,
and the evolution of clusters of galaxies have not been included. Furthermore,
the pointlike model for luminous disks of background galaxies needs to
be improved. Both the detailed theoretical consideration and observational
tests for galaxy-cluster associations will be made soon.

   \acknowledgements
WXP wishes to thank CNRS and K.C. Wong Foundation for financial support.

%

\end{document}